\documentclass[lettersize,journal]{IEEEtran}
\usepackage{amsmath,amsfonts}
\usepackage{algorithmic}
\usepackage{algorithm}
\usepackage{array}
\usepackage[caption=false,font=normalsize,labelfont=sf,textfont=sf]{subfig}
\usepackage{textcomp}
\usepackage{stfloats}
\usepackage{url}
\usepackage{enumerate}
\usepackage{verbatim}
\usepackage{graphicx}
\usepackage{xcolor}
\usepackage{cite}
\usepackage{tabularx,booktabs,textcomp}
\usepackage{amsthm}

\usepackage{amsmath,amssymb,amsthm}

\begin{document}

\title{Causality and Interpretability for Electrical Distribution System faults}

\author{{Karthik Peddi, Sai Ram Aditya Parisineni, Hemanth Macharla, Mayukha Pal*}
\thanks{(Corresponding author: Mayukha Pal)}
\thanks{Mr. Karthik Peddi is a Data Science Research Intern at ABB Ability Innovation Center, Hyderabad 500084, India and also a bachelor student at the Department of Electronics and Communication Engineering, IIT Bhubaneswar, Odisha 752050.}
\thanks{Mr. Sai Ram Aditya Parisineni is a Data Science Research Intern at ABB Ability Innovation Center, Hyderabad 500084, India and also a bachelor student at the Department of Artificial Intelligence, IIT Hyderabad, Telangana 502284.}
\thanks{Mr.Hemanth Macharla is a Data Science Research Intern at ABB Ability Innovation Center, Hyderabad 500084, India, and also a bachelor's student at the Department of Computer Science and Engineering, IIT Bhubaneswar, Odisha 752050.}

\thanks{Dr. Mayukha Pal is with ABB Ability Innovation Center, Hyderabad-500084, IN, working as Global R\&D Leader – Cloud \& Advanced Analytics (e-mail: mayukha.pal@in.abb.com).}}


\maketitle

\begin{abstract} 
Causal analysis helps us understand variables that are responsible for system failures. This improves fault detection and makes system more reliable. In this work, we present a new method that combines causal inference with machine learning to classify faults in electrical distribution systems (EDS) using graph-based models. We first build causal graphs using transfer entropy (TE). Each fault case is represented as a graph, where the nodes are features such as voltage and current, and the edges demonstrate how these features influence each other. Then, the graphs are classified using machine learning and GraphSAGE where the model learns from both the node values and the structure of the graph to predict the type of fault. To make the predictions understandable, we further developed an integrated approach using GNNExplainer and Captum’s Integrated Gradients to highlight the nodes (features) that influences the most on the final prediction. This gives us clear insights into the possible causes of the fault. Our experiments show high accuracy: 99.44\% on the EDS fault dataset, which is better than state of art models. By combining causal graphs with machine learning, our method not only predicts faults accurately but also helps understand their root causes. This makes it a strong and practical tool for improving system reliability.

\end{abstract}


\section{Introduction}
\label{section:Introduction}

\subsection{Background and Motivation}
The integration of causal inference with machine learning \cite{LIME, SHAP} has recently emerged as a promising approach to enhance model interpretability and performance. Often assuming that characteristics are independent predictors, traditional machine learning algorithms ignore the underlying causal links among them. Particularly in complicated, high-dimensional datasets, such oversight produces accurate models but lacks transparency and robustness.

Causal inference aims to uncover and utilize these underlying relationships, offering a deeper understanding of how different variables interact and influence outcomes. Including causal analysis in the classification process helps us boost prediction accuracy and find the key variables that affect these forecasts. This added layer of interpretability is crucial in fields such as healthcare, finance, and social sciences, where understanding the causative factors behind a prediction will inform better decision-making and policy formulation.

This paper proposes a novel approach to classification that combines causal inference with machine learning. We use transfer entropy to construct a graph that captures the causal relationships in the dataset. This graph is then used as input to a GraphSAGE-based prediction model, which leverages the graph structure to improve classification performance. Finally, we perform causal analysis using GNNExplainer and Captum’s Integrated Gradients to identify the most causative variables influencing the predictions.

\subsection{Literature Review}
In revealing intricate temporal relationships, the transformation of time series data into graph structures has become ever more important. Among several techniques, transfer entropy has become a rather effective instrument for identifying linear and nonlinear causal links. Schreiber (2000) first proposed transfer entropy \cite{TE,TE_GAT}, which gauges the directed information flow between two variables, therefore allowing the creation of directed graphs reflecting the fundamental causal structure of the data. Transfer entropy is suitable for a wide range of applications, including financial market analysis, neuroscience, and climatology, unlike conventional correlation-based techniques, since it does not imply linearity and captures more complex connections.

Similarly used in causal graph discovery are various techniques, including Granger causality \cite{Grangercausality,gc_mayukha}, which evaluates the predictability of one time series depending on the historical values of another. Granger causality is limited to linear dependencies, which can be constrictive in complicated systems, even with its simplicity and interpretability. Bayesian networks \cite{Bayseian1, Bayseian2} and their variants, such as Dynamic Bayesian Networks (DBNs) \cite{dynamicbaesiannetwork, Gc_vs_DBN}, provide probabilistic methods for causal discovery, hence allowing uncertainty and including past information. These approaches may have difficulty with scalability in big datasets and frequently call for significant processing resources, though. More advanced methods, such as the Peter and Clark Momentary Conditional Independence (PCMCi) algorithm \cite{PC, PCMCI}, are now being used to solve problems of indirect causality and conditional independence, hence improving the dependability of causal discovery.

The extension of deep learning techniques to non-Euclidean domains has revolutionized the analysis of graph-structured data through Graph Neural Networks (GNNs)\cite{GNN}. In various tasks like node classification, link prediction, and graph classification, GNN variants such as Graph Convolutional Networks (GCNs)\cite{GCN}, Graph Attention Networks (GATs)\cite{GAT ,GATv2}, Graph Isomorphic Networks (GINs)\cite{GIN}, and GraphSAGE\cite{GraphSAGE} have demonstrated exceptional performance. Among these, GraphSAGE has gained attention for its ability to generate node embeddings by sampling and aggregating information from a node's local neighborhood, making it highly effective for scalable learning on large graphs. These GNNs are particularly valuable for analyzing complex graph structures derived from time series data, as they efficiently capture both the topological and feature-based characteristics of graphs.

Understanding and trusting machine learning models is important, especially when the models are complex, like Graph Neural Networks (GNNs). To help with this, we use explanation tools such as GNNExplainer and Captum’s Integrated Gradients. GNNExplainer helps us see which parts of the graph—like specific nodes or edges—are most important for the model’s decision by slightly changing them and seeing the effect. Captum shows how much each feature (like voltage or current) contributes to the final prediction by measuring how the model reacts as we move from a baseline value to the actual input. These tools give us a better understanding of both the structure of the graph and the values of the features, making the model’s decisions easier to explain and trust. Another prominent technique is Local Interpretable Model-agnostic Explanations (LIME), introduced by Ribeiro et al. (2016)\cite{LIME}. LIME simplifies model interpretation by approximating the complex model with a locally interpretable surrogate model near a specific prediction, offering clear local explanations. As a model-agnostic approach, LIME can be applied across various machine-learning models, providing both flexibility and simplicity.

\subsection{Main Contributions and Paper Organisation}
The main contributions of this paper are as follows:

\begin{enumerate}
    \item We build graphs from time-series data using transfer entropy. In these graphs, the nodes represent features like voltage and current, and the edges show how these features affect each other. This gives a strong base for fault classification.
    
    \item We use a machine learning model GraphSAGE to classify these graphs. This model predicts the type of fault by learning from both the node values and the structure of the graph, using node embeddings to make accurate predictions.
    
    \item We use several explanation methods, GNNExplainer and Captum’s Integrated Gradients, to understand why the model makes certain predictions. These tools help us find out which features and nodes are most important for the model's decisions. This makes the model easier to trust and helps us understand the causes of faults more clearly.

\end{enumerate}

By combining causal reasoning with machine learning, this paper aims to build models that are not only accurate but also easy to interpret, helping improve data-driven decision-making. 

This paper is organized as follows: Section \ref{section:Introduction} explains the motivation behind the study and gives an overview of related research. Section \ref{section: References} describes the main concepts used in the paper, such as transfer entropy for building graphs, GraphSAGE for classifying nodes, and GNNExplainer and Captum’s Integrated Gradients for explaining model predictions. Section \ref{Section:Model} Details how we build a separate graph for each data point in the fault case, and how GraphSAGE is set up for classification, and how GNNExplainer and Captum’s Integrated Gradients are used to analyze node importance. Section \ref{section:Results} shows the experimental results using two datasets, including accuracy and explanations from GNNExplainer and Captum’s Integrated Gradients. The paper concludes in Section \ref{section:Conclusion} with a summary of the key results, discusses their importance, and suggests future research directions.

\section{Materials and Methods}
\label{section: References}
\subsection{Transfer Entropy}

In this section, we will explain the graph construction using the input data in tabular form. Our algorithm works with the help of existing causal discovery algorithms to find the causal effect strengths between the variables. In this paper, we utilized the transfer entropy metric to understand the strength of the causal effect. 

Transfer entropy is an information-theoretic measure of causality, first introduced by Schreiber in 2000\cite{TE, TE_GAT}. For a variable \(\mathbf{X} \in \mathbb{R}^t\), its information entropy is defined as:

\begin{equation}
H(X) = - \sum_{i=1}^n p(x_i) \log p(x_i)
\end{equation}

where \(p(x_i)\) is the probability of occurrence of the value \(x_i\). Information entropy quantifies the amount of information within a variable. A higher value of \(H(X)\) suggests that the variable \(X\) contains more information. Conditional entropy, another key concept in information theory, is defined for two variables \(X\) and \(Y\) as \cite{conditional_entropy}:

\begin{equation}
H(X \mid Y) = - \sum_{y \in Y} p(y) \sum_{x \in X} p(x \mid y) \log p(x \mid y)
\end{equation}

where \(p(y)\) is the probability of \(y\), and \(p(x \mid y)\) is the conditional probability of \(x\) given \(y\).

Now TE of \(Y\) to \(X\) is defined as \cite{TE}
\begin{equation}
\begin{aligned}
T_{Y \rightarrow X} = & - \sum_{x_{t+1}, x_t} p(x_{t+1}, x_t) \log p(x_{t+1} \mid x_t) \\
& + \sum_{x_{t+1}, x_t, y_t} p(x_{t+1}, x_t, y_t) \log p(x_{t+1} \mid x_t, y_t) \\
= & \sum_{x_{t+1}, x_t, y_t} p(x_{t+1}, x_t, y_t) \log \frac{p(x_{t+1} \mid x_t, y_t)}{p(x_{t+1} \mid x_t)} \\
= &  \left( H(X_{t+1} \mid X_t) - H(X_{t+1} \mid X_t, Y_t) \right)
\end{aligned}
\end{equation}

where \(x_t\) and \(y_t\) denote the values the respected variables at time t. \( x_t^{(k)} \) = \([ x_{t}, x_{t-1}, \ldots, x_{t-k+1} ]\) and \( y_t^{(l)} \) = \([ y_{t}, y_{t-1}, \ldots, y_{t-l+1} ]\). TE measures the increase in the information amount of the variable \(X\) when \(Y\) is known to when \(Y\) is unknown. TE represents the direction of information flow, thus characterizing causality. As TE is asymmetric, between \(X\) and \(Y\), the causal relationship can be found using the following equation:
\begin{equation}
T_{X,Y} = T_{X \rightarrow Y} - T_{Y \rightarrow X}
\end{equation}

If \(T_{X, Y}\) is positive, then it means that the variable \(X\) is the cause of variable \(Y\). Otherwise, \(X\) is the consequence of \(Y\). We use the above equation to find the causal relationship among the variables. The element \(a_{ij}\) corresponding to the \(i\)-th row and \(j\)-th column of the causality matrix A of the multivariate time-series data can be formulated as:

\begin{equation}
a_{ij} =
\begin{cases} 
T_{v_i, v_j}, & \text{if } T_{v_i, v_j} > c \\
0, & \text{otherwise}
\end{cases}
\end{equation}
where \( v_i \) and \( v_j \) represent the \( i \)-th and \( j \)-th variables of the multivariate timeseries data. We use a threshold value \( c \) to determine whether the causality is significant. The matrix \( A \) serves as the adjacency matrix for the graph structure of the multivariate time series.

\subsection{Graph SAGE}
GraphSAGE, short for Graph Sample and Aggregate, is a prominent framework introduced by Hamilton, Yin, and Leskovec in 2017\cite{GraphSAGE} for learning inductive representation on large graphs. Unlike transductive methods that require retraining for new nodes, GraphSAGE is designed to generalize to unseen nodes, making it particularly suitable for dynamic and evolving graphs.

GraphSAGE creates node embeddings by sampling and collecting local neighborhood features of a node. Neighborhood sampling starts the process: for every node \(v\), a fixed-size sample of its neighbors is chosen. Layer by layer, iteratively, this sampling generates a multihop neighborhood. A fixed subset \(S(v) \subseteq N(v)\) is sampled mathematically, where \(N(v)\) denotes the neighbors of \(v\). The features of the sampled neighbor are then aggregated using functions such as mean, LSTM, or pooling. Capturing the local structure and features of the graph depends on this aggregation process, so it is crucial to select the aggregation step appropriately. Let \(h_v^{(k)}\) denote the embedding of node \(v\) at layer \(k\). One of the often-used aggregators, the mean aggregator, updates the embedding as follows \cite{GraphSAGE}:
\begin{equation}
h_v^{(k+1)} = \sigma \left( W^{(k)} \cdot \text{mean} \left( \{ h_v^{(k)} \} \cup \{ h_u^{(k)} : u \in S(v) \} \right) \right) 
\end{equation}

Where \(W^{(k)}\) is the weight matrix in layer \(k\), \(\sigma\) is a non-linear activation function (for example, ReLU), and \(\text{mean}\) denotes the mean operation of the elements.

Finally, the node’s embedding is updated by combining its current representation with the aggregated neighborhood features. This update rule can be generalized for different aggregation functions, such as LSTM aggregators or pooling aggregators. For instance, the pooling aggregator applies a pooling operation (e.g., max-pooling) to the neighbor embeddings before applying a fully connected layer.

GraphSAGE's design has various advantages. By restricting the number of neighbors taken into account during every aggregation stage, its sampling approach guarantees scalability to big graphs. Large-scale applications where full-batch training is computationally impractical depend especially on this. GraphSAGE also supports inductive learning so that it is able to generate embeddings for nodes unavailable during the training period. This gives flexibility in managing dynamic graphs, including social networks where connections or new users are always developing. GraphSAGE's integration with several aggregation techniques adds even more flexibility. This lets the framework be adapted for different kinds of graph data and applications. For example, whereas in biological networks, more complicated aggregators like LSTM may better reflect the sequential dependencies between nodes, in social networks, the mean aggregator might efficiently capture the average influence of a user's friends. Recent studies have looked at several ways to extend GraphSAGE and improve its functionality even more. Some methods combine attention mechanisms to dynamically balance the significance of various neighbors. Despite these advancements, the core principle of GraphSAGE, efficiently aggregating local neighborhood information, remains central to its design.

In this paper, we integrated Transfer Entropy with GraphSAGE, which allows us to construct graphs that better reflect the underlying dynamics of the data, providing a richer context for GraphSAGE's aggregation process.

\subsection{Explainability Tools
}
GNNExplainer and Captum’s Integrated Gradients offer effective methods for interpreting complex graph-based machine learning models. These explainers help us understand how each feature or graph component contributes to the model’s predictions, which is essential when applying Graph Neural Networks (GNNs) to real-world problems like fault detection in electrical systems.

GNNExplainer works by identifying the most important parts of a graph that influence the prediction. It does this by selectively masking node features or removing edges and observing how these changes affect the model’s output. The goal is to find a minimal subgraph and a set of characteristics that are most critical to the decision of the model. This gives insight into both the structural and feature-level importance, revealing which nodes and connections the model relied on the most.

Captum’s Integrated Gradients is another technique that focuses on feature attribution. It calculates the importance of each feature by measuring the gradient of the model’s output as the input changes from a baseline (like zero) to the actual value. This method captures how much a feature influences the prediction in a smooth and mathematically consistent way. By summing these contributions across all input steps, we get a clear view of how sensitive the model is to each feature.

One of the key strengths of these methods is that they complement each other. While GNNExplainer helps identify which nodes and edges in the graph structure are most important, Integrated Gradients focuses on how individual node features affect the final prediction. This dual perspective provides a more complete explanation of the model’s behavior, both in terms of graph topology and feature content.

These explainers also offer useful properties for model interpretation. GNNExplainer ensures that the explanations are grounded in the actual graph structure and shows which parts of the graph are most critical for a specific prediction. Integrated Gradients maintains consistency by showing that if a feature’s influence on the model increases, its attribution score also increases, making it a reliable tool for analyzing model behavior over different inputs.

In recent years, these tools have been used in many domains, such as healthcare, finance, and power systems, where understanding model predictions is crucial. Their ability to explain both feature-level and structural behavior has made them valuable in making GNNs more transparent and trustworthy.

In this paper, we apply GNNExplainer and Captum’s Integrated Gradients to analyze predictions made by our GraphSAGE-based classification model. By identifying the most important nodes and features that contribute to fault classification, we gain deeper insight into the underlying causes of faults in electrical distribution systems. This makes our model not only highly accurate but also interpretable and practical for real-world applications where explainability is essential.

\begin{figure*}
  \centering
  \includegraphics[width=7.0 in]{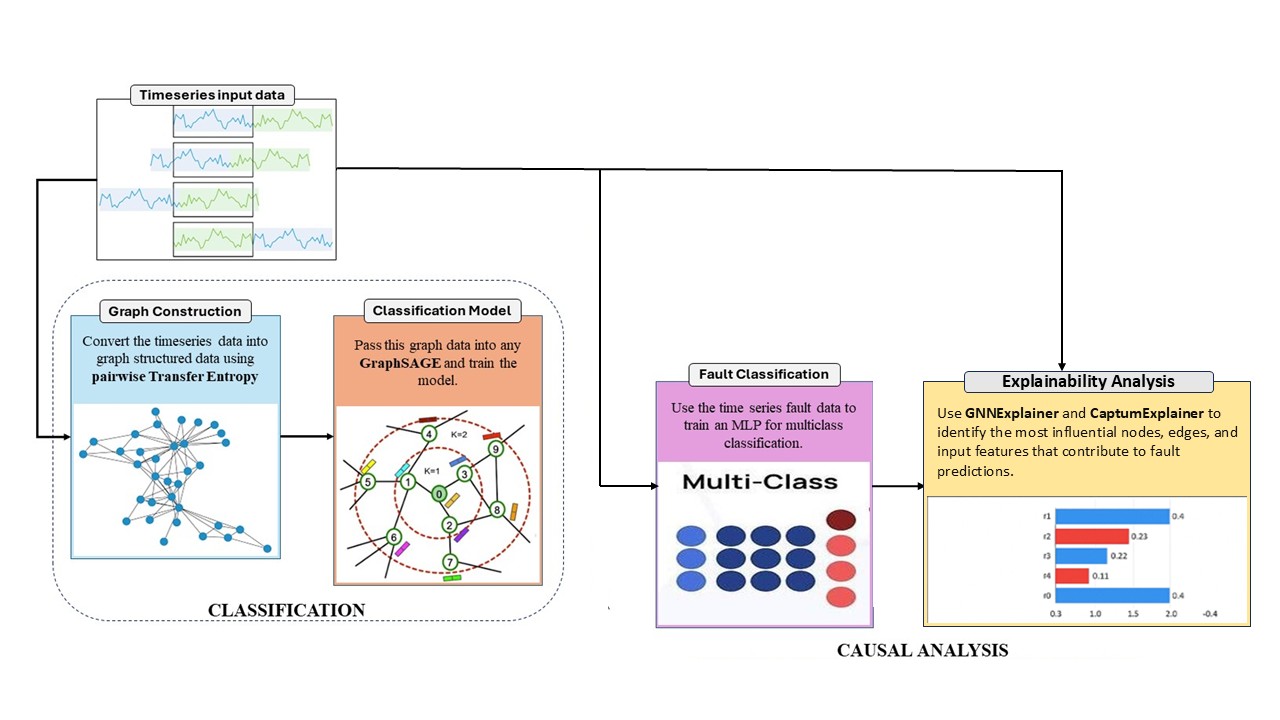}
  \caption{Process flow diagram illustrating the key components of the framework: constructing a graph structure using transfer entropy to capture causal relationships between node features in time series data, applying this structure to a GraphSAGE model for classification, and using Explainability tools for interpreting the model’s predicted outcomes.}
  \label{fig:model_flow}
\end{figure*}

\section{Model Architecture}
\label{Section:Model}

In this section, we outline the model architecture. The model flow diagram in Fig. \ref{fig:model_flow} illustrates the overall process: beginning with preprocessing of time series data, proceeding with graph construction using transfer entropy, and then performing classification with GraphSAGE. The final step involves using GNNExplainer and
Captum’s Integrated Gradients to interpret the model’s predictions. We will provide a detailed discussion of each component below.

\subsection{Graph Construction using Transfer Entropy}
 Transfer entropy measures the directed information flow between variables, making it suitable for capturing causal relationships in complex systems such as networks or time series data.

Each fault case in the dataset is turned into a directed graph, where the nodes represent feature variables (for example, 6 features in the EDS dataset). The node values come from time-series data, which are normalized by subtracting the mean and dividing by the standard deviation. We calculate transfer entropy (TE) between every pair of features using the pyinform library. This gives a TE matrix for each fault instance. The matrix is then normalized, a threshold of 0.2 is applied, and it is turned into a binary adjacency matrix, where directed edges are added only if $TE_{x \ x\rightarrow y} > TE_{y \rightarrow x}$. These edges in the graph show causal relationships between the features. For example, in the EDS dataset, each graph has 6 nodes (representing 3-phase voltages and currents), and the edges are drawn based on the TE values. Note that we do not use edge weights in our graph construction, as the GraphSAGE mode, which we apply for subsequent analysis, does not make use of edge weights. Instead, GraphSAGE focuses on node features and the graph structure to perform learning and make predictions. GraphSAGE was designed to be a general-purpose graph neural network that can aggregate information from neighbors regardless of edge weights. This simplification helps in broad applicability and ease of implementation.

To improve the graph further, we use Direct Transfer Entropy (DTE). This removes indirect causal edges, so that only direct relationships between nodes remain. This makes the graph more suitable for the classification task and also makes sure that the edges in the graph represent direct causal relationships among the nodes. The DTE between two nodes \(x\) and \(y\), assuming an intermediate variable as \(z\), is defined as follows \cite{dte}:

\begin{equation}
DTE_{X \rightarrow Y } = \sum_{y+1, y, z, x} p(y_{t+1}| y_t, z, x) \log \frac{p(y_{t+1} | y_t, z, x)}{p(y_{t+1} | y_t, z)}
\end{equation}

If $ DTE_{X \rightarrow Y} = 0$, then x to y has an indirect causal relationship; otherwise, x and y have a direct causal relationship.

\subsection{Integration with GraphSAGE}
Each causal graph, representing a fault instance, is fed into a GraphSAGE model for graph classification. For each graph, the node features (such as 6001 time points for EDS) and edge connections are used to generate node embeddings through a mean aggregator. These embeddings are then combined using mean pooling to create a single graph-level embedding. This graph-level embedding is passed through linear layers to predict one label per graph—for example, the fault type in the EDS dataset. The model uses two GraphSAGE layers, with the first reducing input dimensions (6001 or 34) to 8, and the second keeping it at 8, followed by linear layers with ReLU activation and dropout to improve generalization. By leveraging the causal structure of the graphs, this approach improves classification accuracy and generalizes well to new, unseen fault scenarios.

\subsection{Analysis Using Explainability Tools}
After obtaining predictions from the GraphSAGE-based model, the next step is to perform causal analysis to identify the most influential variables driving these predictions. This is achieved using two explainability tools, GNNExplainer and Captum’s Integrated Gradients, which help us understand the importance of different parts of the graph in a clear and unified way.

The procedure begins by applying GNNExplainer to the graphs used in our GraphSAGE model. GNNExplainer looks at the graph structure and identifies which nodes and connections (edges) are most important for the model’s predictions by testing how changes to the graph affect the results. For example, it might remove some edges or features and check how much the prediction changes, helping us see which parts of the graph matter most. At the same time, we use Captum’s Integrated Gradients, which focuses on the node features (like voltage or current values in the EDS dataset). This method measures how much each feature contributes to the prediction by looking at how the model’s output changes when we adjust the feature values step by step.

We then combine the results from GNNExplainer and Integrated Gradients to get a complete picture. For each graph, both tools give us importance scores for the 6 nodes (representing the 3-phase voltages and currents in the EDS dataset). We adjust these scores to a common range (between 0 and 1) and take their average, creating a single combined score for each node. This combined score tells us which nodes are the most influential in driving the model’s predictions for each graph.

Using this approach offers several benefits for understanding the model’s decisions, and by integrating GNNExplainer and Integrated Gradients into our methodology, we pinpoint the key variables that drive the classification outcomes, offering valuable insights into the graph structure and feature importance. This step ensures that our approach not only improves prediction accuracy but also provides a clear and reliable way to understand the factors influencing the model’s decisions.

\section{Results and Discussion}
\label{section:Results}

To evaluate the effectiveness of our proposed approach, we conducted experiments on one dataset: the generated electrical multiclass fault classification dataset. The goal was to demonstrate the utility of transfer entropy in constructing meaningful graph structures, assess the performance of the GraphSAGE model in these graph-based classification tasks, and evaluate the interpretability of the model's predictions using GNNExplainer and Captum’s Integrated Gradients.

\begin{figure}[H]
  \centering
  \includegraphics[width=2.7 in]{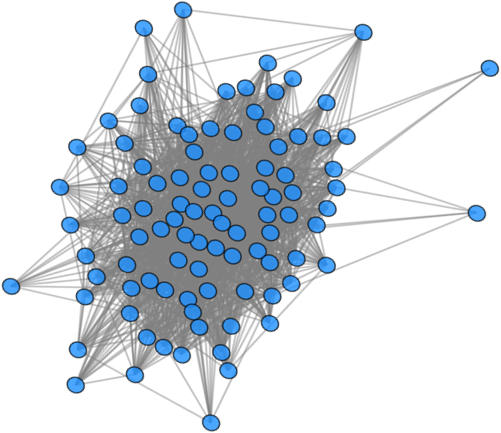}
  \caption{Graph structure constructed by applying transfer entropy to identify causal links between node features in time series data. }
  \label{fig:graph_network}
\end{figure}

\begin{figure}[h]
  \centering
  \includegraphics[width=3.4in]{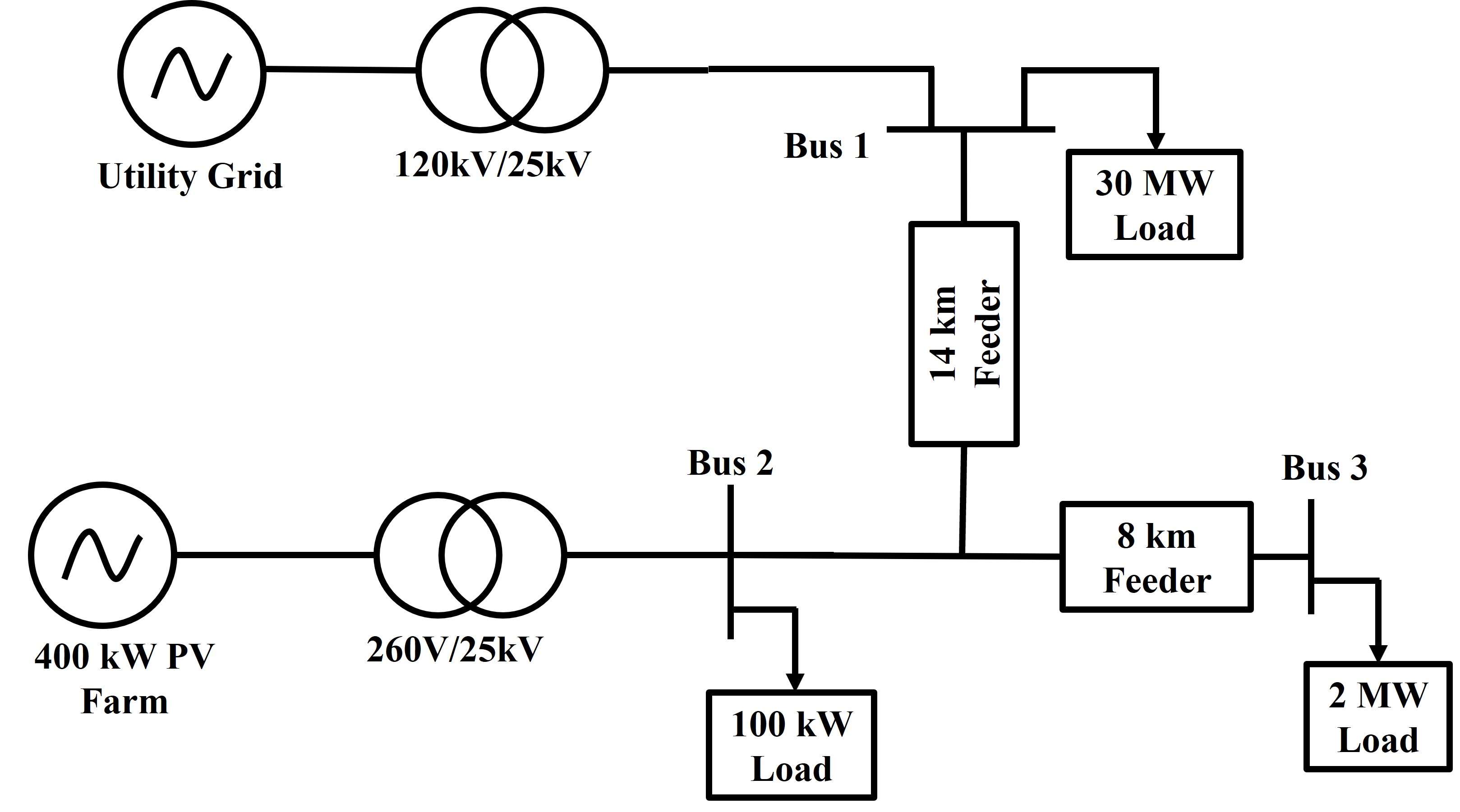}
  \caption{Single line diagram of the considered EDS for the generated dataset.}
  \label{fig:system}
\end{figure}

\subsection{Generated Dataset}
\subsubsection{Details for Considered Dataset}
The electrical distribution system used in this study is shown in Fig. \ref{fig:system}. It has two distribution feeders that are 8 km and 14 km long. A 400 kW solar PV farm is connected to Bus 2, while the main power supply from the grid is connected to Bus 1. Each of the three buses in the system has three loads connected. All measurement data is collected at Bus 2. We simulated different fault scenarios to create a dataset with 9 different fault types. Each fault instance includes 6 time-series features: 3-phase voltages and 3-phase currents, which show how the electrical signals change over time. The fault names describe which phases are affected. For example, ‘AB’ means faults in phases A and B; ‘ABG’ means faults in phases A, B, and ground; ‘AG’ means faults in phases A and ground; ‘BC’ means faults in phases B and C; ‘BCG’ means faults in phases B, C, and ground; ‘BG’ means faults in phases B and ground; ‘CA’ means faults in phases C and A; ‘CAG’ means faults in phases C, A, and ground; ‘CG’ means faults in phases C and ground.

\begin{table}[htbp]
\centering
\caption{Model Parameters}
\begin{tabular}{lcc}
\toprule
Parameters & Values \\
\midrule
Test split & 20\%  \\
Hidden Layer dimension & 256 (first layer), 128 (second layer)	 \\
Optimizer & Adam\\
Learning Rate & 0.00001  \\
No of Epochs & 100	\\
Dropout & 0 \\
Batch Size & 1 \\
\bottomrule
\end{tabular}
\label{tab:model_parameters}
\end{table}

\subsubsection{Results}
\begin{figure}
  \centering
  \includegraphics[width=3.1in]{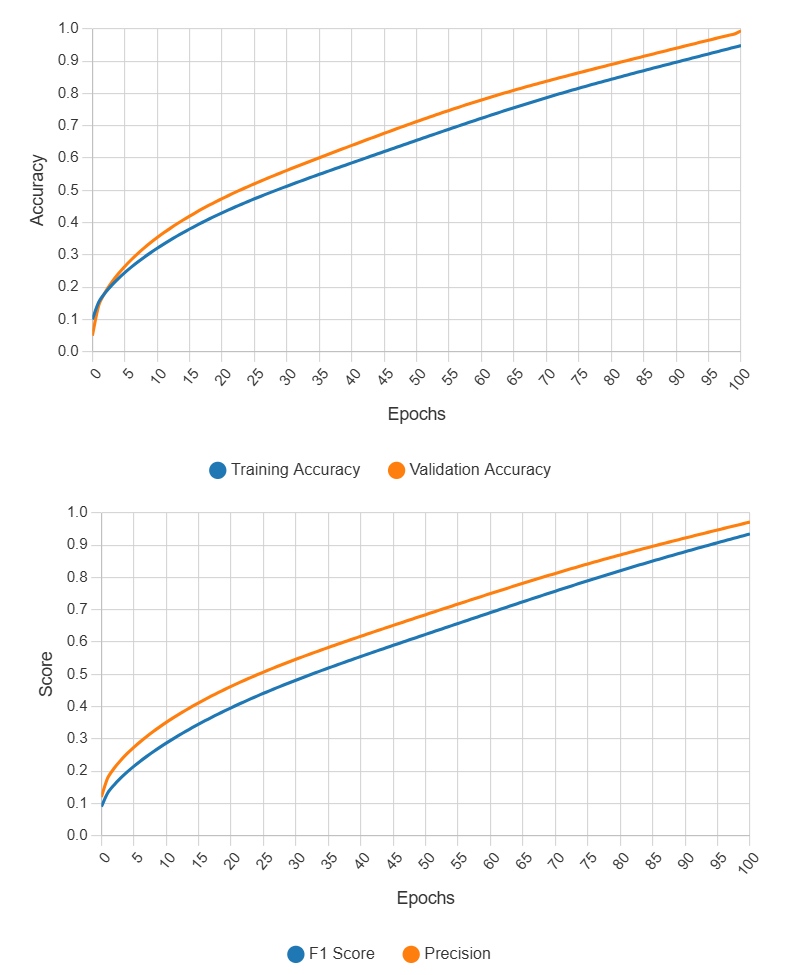}
  \caption{Plot showing the train accuracy, validation accuracy, precision, and F1 score over epochs using the GraphSAGE classification model on the Electrical Multiclass Fault classification dataset.}
  \label{fig:acc_prec_electrical}
\end{figure}

We began by applying transfer entropy to the time series data to construct a directed graph, capturing the causal relationships between the features. The graph structure corresponding to this dataset is shown in Fig. \ref{fig:graph_network}. This graph structure was then fed into the GraphSAGE model for classification. To check the model's performance, we used several evaluation metrics, including accuracy, precision, recall, and F1-score.
We used this GraphSAGE model with the hyperparameters listed in Table \ref{tab:model_parameters}. On the test data, the model achieved an average precision of 99.44\% and a precision of about 99.42\%. Figure \ref{fig:acc_prec_electrical} shows how accuracy, precision, and F1 score changed over the training epochs. Since our TE with the GraphSAGE framework already showed strong performance, we now focus on explaining the classification results on this dataset using GNNExplainer and CaptumExplainer.

\medskip
For multiclass classification with GraphSAGE, we focus on explaining the model’s predictions for different fault types by finding which nodes have the biggest influence on the decisions. We use two explanation methods: GNNExplainer and Captum’s Integrated Gradients, which together give a fuller picture of feature importance. GNNExplainer works by changing parts of the graph (like hiding some edges or features) and seeing how that affects the prediction. This helps highlight which nodes and edges are most important in the graph’s structure. Integrated Gradients, from Captum, calculates feature importance by looking at how the model’s output changes as the input (node features and edges) moves from a baseline to the actual data, showing how much each node feature contributes to the final prediction.

\medskip

We propose an embedding model that combines the strengths of GNNExplainer and Integrated Gradients to improve the reliability of our interpretability analysis.

\medskip

For each graph \( G_i \) (where \( i \in \{0, \dots, 899\} \)), both explainers give importance scores for the 6 nodes, showing how much each node influences the GraphSAGE model’s prediction.

\begin{itemize}
    \item \textbf{GNNExplainer} focuses on structural connections by finding important subgraphs through changing edges and features.
    \item \textbf{Integrated Gradients} focuses on feature-level effects by integrating gradients of the model’s output with respect to input features from a baseline to the actual input.
\end{itemize}

\medskip

To combine these insights, we normalize the importance scores from both methods to a common scale between 0 and 1, then calculate their average to get a combined importance score for each node. Let \( S_{\text{GNN}}(v_j, G_i) \) and \( S_{\text{IG}}(v_j, G_i) \) be the importance scores for node \( v_j \) (where \( j \in \{0, \dots, 5\} \)) in graph \( G_i \) from GNNExplainer and Integrated Gradients, respectively.

The normalized scores are:

\[
S'_{\text{GNN}}(v_j, G_i) = \frac{S_{\text{GNN}}(v_j, G_i) - \min_k S_{\text{GNN}}(v_k, G_i)}{\max_k S_{\text{GNN}}(v_k, G_i) - \min_k S_{\text{GNN}}(v_k, G_i)},
\]

\[
S'_{\text{IG}}(v_j, G_i) = \frac{S_{\text{IG}}(v_j, G_i) - \min_k S_{\text{IG}}(v_k, G_i)}{\max_k S_{\text{IG}}(v_k, G_i) - \min_k S_{\text{IG}}(v_k, G_i)},
\]

Where the minimum and maximum values are taken over all nodes \( v_k \) in the graph \( G_i \).

The combined importance score is then:

\[
S_{\text{combined}}(v_j, G_i) = \frac{1}{2} \left( S'_{\text{GNN}}(v_j, G_i) + S'_{\text{IG}}(v_j, G_i) \right).
\]

This averaging brings together the complementary views into a single score that balances both structural and feature-based influences.

Using this combined score, we rank the top 6 nodes for each graph in a fault class. For the 100 graphs of each class, we count how often each node appears in each rank (from 0th to 5th), helping us identify the key nodes driving fault predictions (for example, nodes 0 and 5 often rank highest).

The nodes correspond to:

\begin{itemize}
  \item 0: \( V_A \) (Phase A voltage)  
  \item 1: \( V_B \) (Phase B voltage)  
  \item 2: \( V_C \) (Phase C voltage)  
  \item 3: \( I_A \) (Phase A current)  
  \item 4: \( I_B \) (Phase B current)  
  \item 5: \( I_C \) (Phase C current)  
\end{itemize}

\medskip

\noindent
This method improves understanding by identifying important features in the graph, like specific voltage or current readings, that strongly affect the model’s choice to label an instance as a fault. Our combined use of GNNExplainer and Integrated Gradients gives a stronger explanation designed for graph data, showing both structural and feature-level effects in multiclass fault classification. This process looks at the causes of faults and how different variables influence the results.

\begin{figure}
  \centering
  \includegraphics[width=3in]{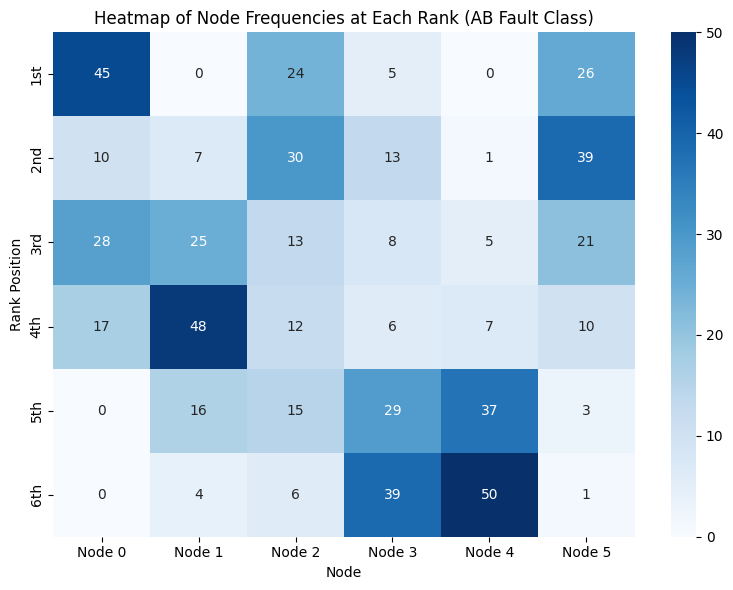}
  \caption{A Heatmap for AB fault. This figure accurately explains the top 2 most impacting Phase nodes that contribute majourly to AB fault. The top important nodes form the heat map are  Phase A voltage ($ V_A $) and Phase C current ($ I_C $).}
  \label{fig:ABF1}
\end{figure}

 \begin{figure}
  \centering
  \includegraphics[width=3.4 in]{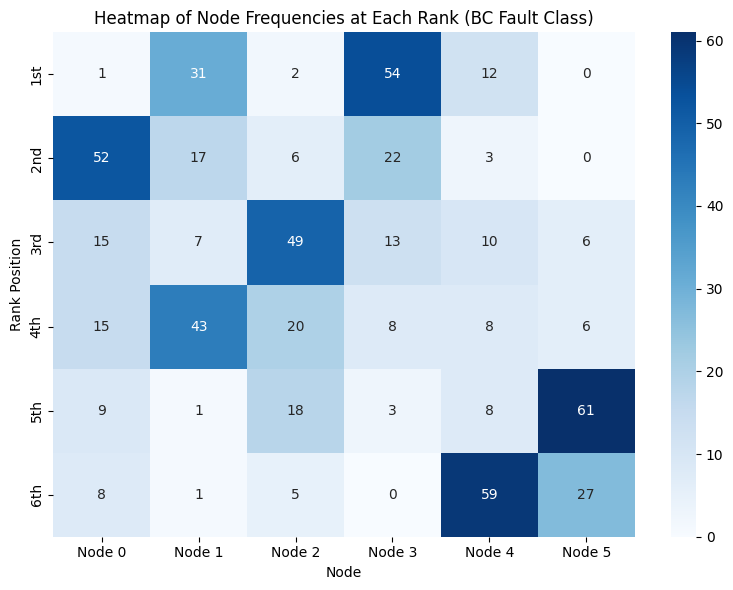}
  \caption{A Heatmap for BC fault. This figure accurately explains the top 2 most impacting Phase nodes that contribute majourly to the BC fault.
}
  \label{fig:BCG_all}
\end{figure}

 \begin{figure}
  \centering
  \includegraphics[width=3.4 in]{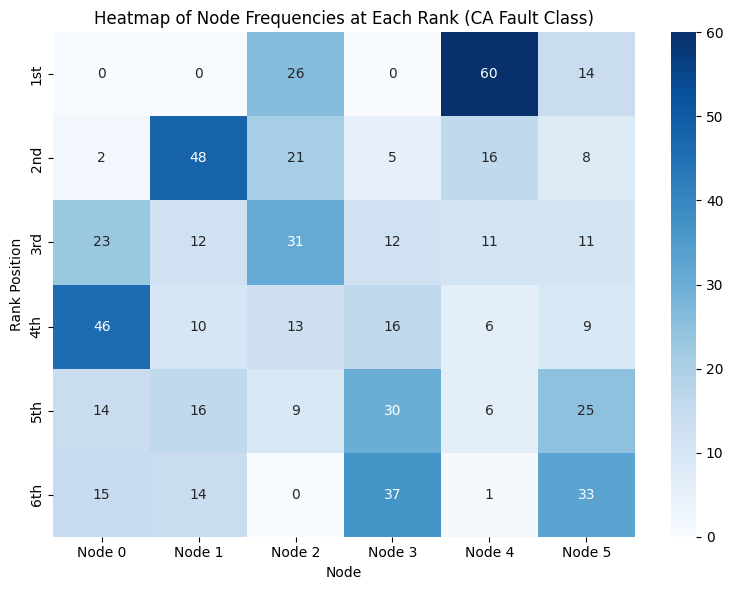}
  \caption{A Heatmap for CA fault. This figure accurately explains the top 2 most impacting Phase nodes that contribute majourly to the CA fault.}
  \label{fig:CG_all}
\end{figure}

\medskip

We applied the explained methods to all fault types. For the AB fault, Fig. \ref{fig:ABF1} shows a heatmap based on 100 AB-fault cases, indicating that Phase A voltage ($V_A$) and Phase C current ($I_C$) are the most important features. The importance of $V_A$ fits with the AB fault since it directly affects Phase A, where voltage changes occur due to the short circuit between Phases A and B. The high importance of $I_C$, which is not directly part of the faulted phases, may be due to interactions in the three-phase system, possibly fault currents causing changes in Phase C or measurement relationships in the data. This shows how GNNExplainer and Integrated Gradients can reveal unexpected effects and give a deeper understanding of the fault’s influence on the system.

\medskip

For the BC fault, Fig. \ref{fig:BCG_all} points out Phase A current ($I_A$) and Phase A voltage ($V_A$) as the most important features.
Similarly, for the CA fault, Fig. \ref{fig:CG_all} shows that Phase B current ($I_B$) and Phase B voltage ($V_B$) are the most important features.

\section{Conclusion}
\label{section:Conclusion}
In this paper, we presented a framework that combines causal graphs with a GraphSAGE model for graph classification, tested on a custom electrical distribution system (EDS) fault dataset. For the EDS dataset, we created 900 graphs with 6 nodes each, representing three-phase voltages and currents across 9 fault classes. The edges in these graphs were also defined by transfer entropy. Our GraphSAGE model showed strong results, achieving 99.44\% on the EDS fault data, outperforming other baseline methods.

\medskip

Using GNNExplainer and CaptumExplainer, we identified important nodes in the graphs, such as nodes 0 and 5 for AB faults in the EDS dataset, which helped us understand the key causal factors behind the predictions. By integrating causal modeling through transfer entropy with graph-based machine learning, our framework improves both the accuracy and explainability of fault classification, supporting better decision-making in electrical systems.

\medskip

\noindent
Overall, our method effectively combines causal analysis with advanced graph learning, offering a strong approach for accurate and understandable data analysis. The high performance and clear explanations provided by our model show the importance of adding causal knowledge to machine learning, helping enable more trustworthy, data-based decisions.

\bibliographystyle{IEEEtran}
\bibliography{main}

\begin{thebibliography}{10}
\providecommand{\url}[1]{#1}
\csname url@samestyle\endcsname
\providecommand{\newblock}{\relax}
\providecommand{\bibinfo}[2]{#2}
\providecommand{\BIBentrySTDinterwordspacing}{\spaceskip=0pt\relax}
\providecommand{\BIBentryALTinterwordstretchfactor}{4}
\providecommand{\BIBentryALTinterwordspacing}{\spaceskip=\fontdimen2\font plus
\BIBentryALTinterwordstretchfactor\fontdimen3\font minus \fontdimen4\font\relax}
\providecommand{\BIBforeignlanguage}[2]{{%
\expandafter\ifx\csname l@#1\endcsname\relax
\typeout{** WARNING: IEEEtran.bst: No hyphenation pattern has been}%
\typeout{** loaded for the language `#1'. Using the pattern for}%
\typeout{** the default language instead.}%
\else
\language=\csname l@#1\endcsname
\fi
#2}}
\providecommand{\BIBdecl}{\relax}
\BIBdecl

\bibitem{LIME}
M.~T. Ribeiro, S.~Singh, and C.~Guestrin, ``" why should i trust you?" explaining the predictions of any classifier,'' in \emph{Proceedings of the 22nd ACM SIGKDD international conference on knowledge discovery and data mining}, 2016, pp. 1135--1144.

\bibitem{SHAP}
S.~M. Lundberg and S.-I. Lee, ``A unified approach to interpreting model predictions,'' \emph{Advances in neural information processing systems}, vol.~30, 2017.

\bibitem{TE}
T.~Schreiber, ``Measuring information transfer,'' \emph{Phys. Rev. Lett.}, vol.~85, pp. 461--464, Jul 2000.

\bibitem{TE_GAT}
S.~Liang, D.~Pi, and X.~Zhang, ``Anomaly detection model for large-scale industrial systems using transfer entropy and graph attention network,'' \emph{Measurement Science and Technology}, vol.~35, no.~9, p. 095104, jun 2024.

\bibitem{Grangercausality}
C.~W. Granger, ``Investigating causal relations by econometric models and cross-spectral methods,'' \emph{Econometrica: journal of the Econometric Society}, pp. 424--438, 1969.

\bibitem{gc_mayukha}
D.~Dwivedi, D.~M. Reddy, P.~K. Yemula, and M.~Pal, ``Identification of critical nodes using granger causality for strengthening network resilience in electrical distribution system,'' in \emph{International Conference on Electrical and Electronics Engineering}.\hskip 1em plus 0.5em minus 0.4em\relax Springer, 2023, pp. 49--60.

\bibitem{Bayseian1}
I.~Ben-Gal, ``Bayesian networks,'' \emph{Encyclopedia of statistics in quality and reliability}, 2008.

\bibitem{Bayseian2}
D.~Heckerman, ``A bayesian approach to learning causal networks,'' \emph{arXiv preprint arXiv:1302.4958}, 2013.

\bibitem{dynamicbaesiannetwork}
K.~P. Murphy \emph{et~al.}, ``Dynamic bayesian networks,'' \emph{Probabilistic Graphical Models, M. Jordan}, vol.~7, p. 431, 2002.

\bibitem{Gc_vs_DBN}
C.~Zou and J.~Feng, ``Granger causality vs. dynamic bayesian network inference: a comparative study,'' \emph{BMC bioinformatics}, vol.~10, pp. 1--17, 2009.

\bibitem{PC}
P.~Spirtes and C.~Glymour, ``An algorithm for fast recovery of sparse causal graphs,'' \emph{Social science computer review}, vol.~9, no.~1, pp. 62--72, 1991.

\bibitem{PCMCI}
C.~Krich, J.~Runge, D.~G. Miralles, M.~Migliavacca, O.~Perez-Priego, T.~El-Madany, A.~Carrara, and M.~D. Mahecha, ``Estimating causal networks in biosphere--atmosphere interaction with the pcmci approach,'' \emph{Biogeosciences}, vol.~17, no.~4, pp. 1033--1061, 2020.

\bibitem{GNN}
F.~Scarselli, M.~Gori, A.~C. Tsoi, M.~Hagenbuchner, and G.~Monfardini, ``The graph neural network model,'' \emph{IEEE transactions on neural networks}, vol.~20, no.~1, pp. 61--80, 2008.

\bibitem{GCN}
T.~N. Kipf and M.~Welling, ``Semi-supervised classification with graph convolutional networks,'' \emph{arXiv preprint arXiv:1609.02907}, 2016.

\bibitem{GAT}
P.~Velickovic, G.~Cucurull, A.~Casanova, A.~Romero, P.~Lio, Y.~Bengio \emph{et~al.}, ``Graph attention networks,'' \emph{stat}, vol. 1050, no.~20, pp. 10--48\,550, 2017.

\bibitem{GATv2}
S.~Brody, U.~Alon, and E.~Yahav, ``How attentive are graph attention networks?'' \emph{arXiv preprint arXiv:2105.14491}, 2021.

\bibitem{GIN}
K.~Xu, W.~Hu, J.~Leskovec, and S.~Jegelka, ``How powerful are graph neural networks?'' \emph{arXiv preprint arXiv:1810.00826}, 2018.

\bibitem{GraphSAGE}
W.~Hamilton, Z.~Ying, and J.~Leskovec, ``Inductive representation learning on large graphs,'' \emph{Advances in neural information processing systems}, vol.~30, 2017.

\bibitem{conditional_entropy}
C.~E. Shannon, ``A mathematical theory of communication,'' \emph{The Bell system technical journal}, vol.~27, no.~3, pp. 379--423, 1948.

\bibitem{dte}
P.~Duan, F.~Yang, T.~Chen, and S.~L. Shah, ``Direct causality detection via the transfer entropy approach,'' \emph{IEEE transactions on control systems technology}, vol.~21, no.~6, pp. 2052--2066, 2013.

\end{thebibliography}

\end{document}